\def\year{2022}\relax
\documentclass[letterpaper]{article} 
\usepackage{aaai22}  
\usepackage{times}  
\usepackage{helvet}  
\usepackage{courier}  
\usepackage[hyphens]{url}  
\usepackage{graphicx} 
\usepackage{natbib}  
\usepackage{caption} 
\DeclareCaptionStyle{ruled}{labelfont=normalfont,labelsep=colon,strut=off} 
\usepackage{algorithm}
\usepackage{algorithmic}
\usepackage{times}
\usepackage{nameref}

\usepackage{newfloat}
\usepackage{listings}
\floatstyle{ruled}
\newfloat{listing}{tb}{lst}{}
\floatname{listing}{Listing}
\usepackage{mathtools}
\usepackage{tabularx}
\usepackage{booktabs}
\usepackage{comment}

\title{IB-U-Nets: Improving medical image segmentation tasks with 3D Inductive Biased kernels }
\author {
    Shrajan Bhandary\textsuperscript{\rm 1},
     Zahra Babaiee\textsuperscript{\rm 1},
     Dejan Kostyszyn\textsuperscript{\rm 2, \rm 3, \rm 4},
     Tobias Fechter\textsuperscript{\rm 2, \rm 3, \rm 4},
     Constantinos Zamboglou\textsuperscript{\rm 2, \rm 3, \rm 4, \rm 5},
     Anca-Ligia Grosu\textsuperscript{\rm 2, \rm 3, \rm 4},
     Radu Grosu\textsuperscript{\rm 1}
}
\affiliations {
    \textsuperscript{\rm 1} Technische Universität Wien, Vienna, Austria\\
    \textsuperscript{\rm 2} Division of Medical Physics, Department of Radiation Oncology, Medical Center University of Freiburg, Germany  \\
    \textsuperscript{\rm 3} Faculty of Medicine, University of Freiburg, Germany  \\
    \textsuperscript{\rm 4} German Cancer Consortium (DKTK), Partner Site Freiburg, Germany \\
    \textsuperscript{\rm 5} German Oncology Center, European University, Limassol, Cyprus \\
    shrajan.bhandary@tuwien.ac.at, zahra.babaiee@tuwien.ac.at,
    dejan.kostyszyn@uniklinik-freiburg.de, 
    tobias.fechter@uniklinik-freiburg.de, 
    constantinos.zamboglou@uniklinik-freiburg.de, 
    anca.grosu@uniklinik-freiburg.de, 
    radu.grosu@tuwien.ac.at
}

\usepackage{bibentry}

\begin{document}

\maketitle

\begin{abstract}
   Despite the success of convolutional neural networks for 3D medical-image segmentation, the architectures currently used are still not robust enough to the protocols of different scanners, and the variety of image properties they produce. Moreover, access to large-scale datasets with annotated regions of interest is scarce, and obtaining good results is thus difficult. To overcome these challenges, we introduce IB-U-Nets, a novel architecture with inductive bias, inspired by the visual processing in vertebrates. With the 3D U-Net as the base, we add two 3D residual components to the second encoder blocks. They provide an inductive bias, helping U-Nets to segment anatomical structures from 3D images with increased robustness and accuracy. We compared IB-U-Nets with state-of-the-art 3D U-Nets on multiple modalities and organs, such as the prostate and spleen, using the same training and testing pipeline, including data processing, augmentation and cross-validation. Our results demonstrate the superior robustness and accuracy of IB-U-Nets, especially on small datasets, as is typically the case in medical-image analysis. IB-U-Nets source code and models are publicly available.
\end{abstract}

\section{Introduction}
\label{Introduction}
Manual segmentation of organs from medical images is an essential tool for diagnosis, treatment, and progression monitoring of several diseases~\cite{goldenberg_NATURE_2019, thno61207, 3d_survey_singh_2020, MDC_2022, nnUNet_2020}. Unfortunately, manual segmentation is time-consuming and underlies significant interobserver heterogeneity~\cite{rischke_2013, steenbergen_2015}. Fortunately, recent advances in deep learning (DL) hold a great promise for automatic segmentation. Since the success of AlexNet~\cite{alexnet_2012}, the biomedical-imaging community has been investigating various convolutional neural-network  architectures (CNN) to enhance cancer-detection efficiency~\cite{3d_survey_singh_2020}. In particular, U-Nets yielded most accurate segmentation on medical datasets~\cite{ronneberger_u_net_2015}, first in 2D and later on in 3D medical-imaging scenarios~\cite{cicek_3d_2016}. However, the U-Nets currently in use are still not sufficiently robust to images obtained from different scanners, to shape variability, and to poor tissue contrast of some organs~\cite{gillespie2020deep}. 
Moreover, the segmentation accuracy of U-Nets still leaves room for improvement, especially in small-sized 3D datasets~\cite{litjens_2017, nnUNet_2020}.

\begin{figure}[t]
\centering
\includegraphics[width=\linewidth]{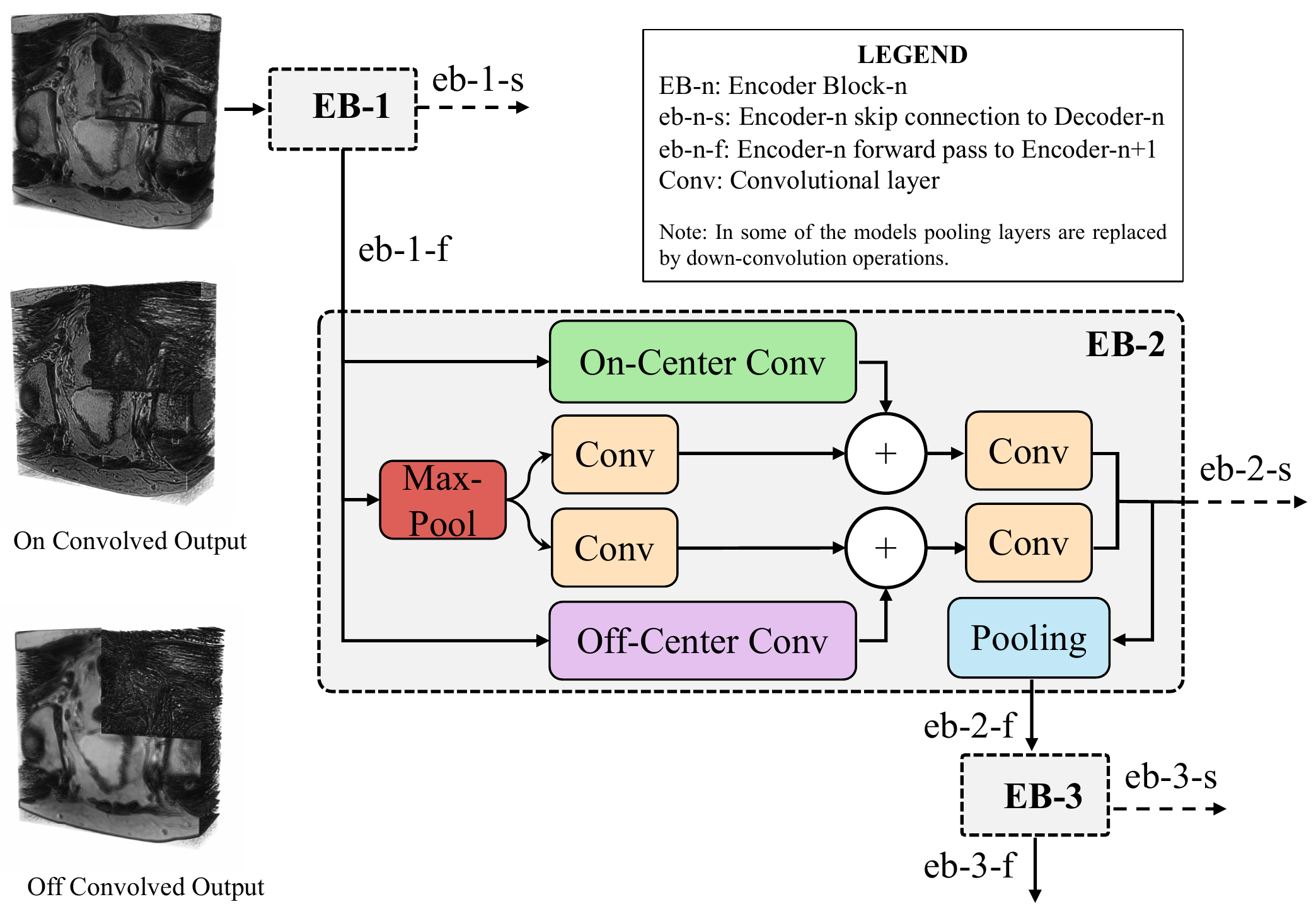}
\caption{Extending U-Net variants with two 3D inductive biases, On and Off center-surround convolution respectively, enhances their segmentation accuracy and robustness to distribution shifts.}
\label{fig:oocsArchitecture}
\vspace{-2mm}
\end{figure}

\begin{figure*}[t]
\centering
\includegraphics[width=\linewidth]{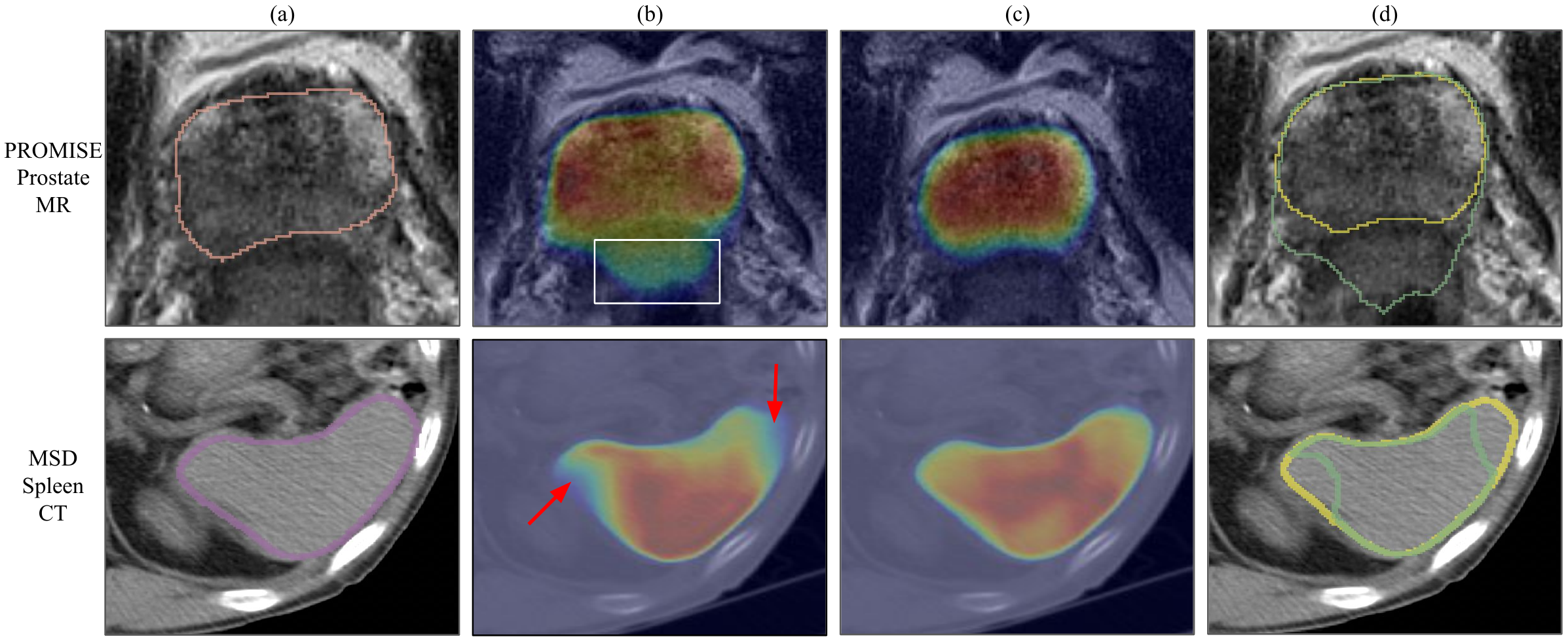}
\caption{Comparison of the attention maps between the nnU-Net and IB-nnU-Net for volumes with a significant difference in accuracy. (a) The raw MR and CT images with annotated regions of interest – in PROMISE-prostate the red contour is the prostate, and in MSD-spleen the purple contour is the spleen. (b) The attention maps of nnU-Net for the given input volume. The white box in the attention map of nnU-Net for the PROMISE-prostate input signifies the area that should not have been activated. Whereas, the red arrows in the attention map for the MSD-spleen image denote sections missed by the model. (c) The attention maps of IB-nnU-Net for the given input volume. (d) Final predictions made by the networks for each task; green and yellow contours correspond to outputs of nnU-Net and IB-nnU-Net, respectively.}
\label{fig:attention_maps}
\end{figure*}

In this paper we address the above limitations, by enhancing state-of-the-art (SOTA) 3D U-Nets, with an inductive bias (IB) inspired by the retinal-processing pathways of vertebrates. In particular, we add two 3D residual components (an on- and an off-center-surround convolution, with predefined weights) to the second encoder blocks of U-Nets (in parallel to the max-pooling component), as depicted in Figure~\ref{fig:oocsArchitecture}. The IB provided by these components, helps the U-Nets to scrutinize and delineate anatomical structures of 3D images with increased precision, as shown in Figure~\ref{fig:attention_maps}.  

For a comprehensive analysis, we implemented several alternative architectures with IB extensions and evaluated robustness against changes in parameter values. 
To demonstrate the general applicability of our method we performed experiments on 3D medical multi modal image datasets containing unique organs. A limiting factor in multiple clinical studies is the low number of datasets available for training. To address this issue, we conducted multiple experiments with varying number of training datasets to test clinical suitability. 
We performed our experiments on two prostate-contour-annotated datasets: PROMISE-12 and MSD-prostate. PROMISE-12 (prostate MRI segmentation 2012) challenge~\cite{LITJENS2014359} and MSD-prostate (medical segmentation decathlon)~\cite{MDC_2019, MDC_2021, MDC_2022} are two of the most widely used and publicly available datasets. Additionally, we carried out experiments with the spleen dataset from medical segmentation decathlon (MSD-spleen)~\cite{MDC_2019, MDC_2021, MDC_2022}. We created a framework, similar to the design principles of other popular frameworks such as MONAI~\cite{monai} and nnU-Net~\cite{nnUNet_2020}, to objectively compare and evaluate different U-Net variants. We also extended the U-Net architecture of the nnU-Net with the IB kernels to examine their effectiveness on a strong and versatile medical segmentation baseline. In particular, we provide our results for U-Nets (nnU-Net), Attention-U-Nets~\cite{attention_unet_2018} and SegResNets~\cite{litjens_2017}, and their IB-extensions, as these networks had the highest accuracy in our experiments. The 3D IB filters are modular, independent, and easily deployable to any existing U-Net architecture. Our extension of SOTA U-Nets shows an increased performance in automatic organ segmentation from 3D MR and CT images and superior robustness for small datasets.

\textbf{In summary our paper:} (1)~Introduces novel 3D inductive biase (IB) and shows how to use them in U-Net variations for optimal performance. (2)~Extends nnU-Nets, Attention-U-Nets and SegResNets with the above IB and shows superior accuracy and robustness on datasets of multiple organs with different modalities.


\section{Related Work}
\label{related_work}

\subsection{U-Net Variations.}
This paper is mainly concerned with U-Nets~\cite{ronneberger_u_net_2015,cicek_3d_2016} and their variations~\cite{nnUNet_2020,attention_unet_2018,segresnet2018,vnet_2016}, as they are the most popular biomedical-segmentation architecture in recent times. U-Nets have an encoder-decoder structure, where the encoder acts as a classifier identifying key features. In contrast, the decoder reconstructs the image from its low-dimensional discriminative features to its original pixel-level space. One of the variants named V-Net~\cite{vnet_2016} replaces the U-Nets pooling layers with down/up convolutions for smaller memory footprint during training and introduce skip connections within every encoder block to improve convergence and learn residual functions. An alternative called Attention-U-Net~\cite{attention_unet_2018} employs an attention-gate module to automatically learn to emphasize target structures of varying shapes and sizes. 

SegResNet ~\cite{segresnet2018} is another variation of the U-Net that has a segmentation architecture similar to V-Nets, but in addition uses a variational auto-encoder (VAE) branch. A VAE branch reconstructs the input image in order to regularize the shared decoder and impose additional constraints on its layers. On the other hand, the nnU-Net framework~\cite{nnUNet_2020} provides robust, and self-adapting networks based on 2D and 3D vanilla U-Nets, and a U-Net cascade. nnU-Net is the most widely used framework in medical image analysis after it achieved the first rank in multiple challenges. 

\subsection{Medical-Image Segmentation}
\paragraph{Prostate-gland segmentation.}
Over the past decade, significant research has been devoted to automatic segmentation of the prostate from MRIs~\cite{gillespie2020deep}. The original V-Net was trained on the PROMISE-12 dataset, where it achieved top results using dice loss and the model's performance was determined by an overlap-based metric called dice similarity coefficient (DSC) \cite{vnet_2016}. \citet{GHAVAMI2019101558} conducted a survey with four U-Net variants trained on a 3D MR dataset of 232 prostate patient volumes, and the results show that a statistically significant difference exists in the performances of these models. \citet{gcn_2020} investigated an interactive segmentation approach by using a graph CNN to predict the prostate gland from the PROMISE-12 dataset. The nnU-Net framework \cite{nnUNet_2020} achieved superior performance in prostate segmentation from PROMISE-12 and prostate-MSD datasets. 

\paragraph{Spleen segmentation.}
Several studies have applied DL approaches to segment spleen from medical images \cite{ALTINI202230}. \citet{Gibson2018} proposed an encoder-decoder model called Dense V-Network for segmenting eight organs, including the spleen. The architecture of the Dense V-Network is similar to the U-Net, with the exception that they used dense blocks during the encoding path and bilinear upsampling in the decoding path. The nnU-Net framework \cite{nnUNet_2020} again secured the top place for segmenting spleen from the MSD dataset. Inspired by the success of transformers for natural language processing, \citet{hatamizadeh2022unetr} proposed a new architecture called UNEt TRansformers (UNETR). While retaining the shape of the U-Net and without relying on convolutional layers for feature extraction, the UNETR uses a pure vision transformer as its encoder. The UNETR achieved good results on multi-organ segmentations, including the MSD-spleen segmentation task. 

Based on previous works, it is evident that CNN based DL algorithms such as U-Nets are efficient in automatically segmenting organs from medical images. 

\subsection{Biologically Inspired Architectures}
\paragraph{Vision models.}  Initially, these models were inspired by neuro\-science and psychology. In the meantime, both neuroscience and artificial intelligence (AI), particularly computer vision, have enormously progressed. However, most of today's neural networks remained only loosely inspired by the visual system. The interaction between the fields has become less common, compared to the early age of AI, despite the importance of neuroscience in generating insightful ideas for AI~\cite{hassibis, HASSABIS2017245}. One line of inquiry remained in the search for neuroscience-inspired enhancements of CNN architectures to increase their interpretability and robustness~\cite{nayebiNeurips2018, li2019learning, Dapello2020, pmlr-v139-babaiee21a}. In~\cite{nayebiNeurips2018}, local recurrence within cortical areas and long-range feedback from downstream areas to upstream areas are exploited in the design of novel recurrent CNNs, to increase interpretability and improve classification performance. 

\citet{li2019learning} proposed a new regularization technique of CNNs, which uses large-scale neuroscience data to learn more robust neural features in terms of representational similarity. In \citet{Dapello2020}, a CNN architecture whose hidden layers better match with the primate primary visual cortex is proposed, which is more robust to adversarial attacks, such as imperceptibly small and explicitly crafted perturbations. Finally, the authors of \citet{pmlr-v139-babaiee21a} added 2D On and Off center-surround parallel pathways present in the retina of vertebrates to CNNs and improved the robustness of image classification. This paper draws inspiration from U-Net variations and \citet{pmlr-v139-babaiee21a} to design a medical-image segmentation architecture that can learn accurate and robust prostate segmentation from small 3D medical datasets.

\begin{figure}[t]
\centering
\includegraphics[width=\linewidth]{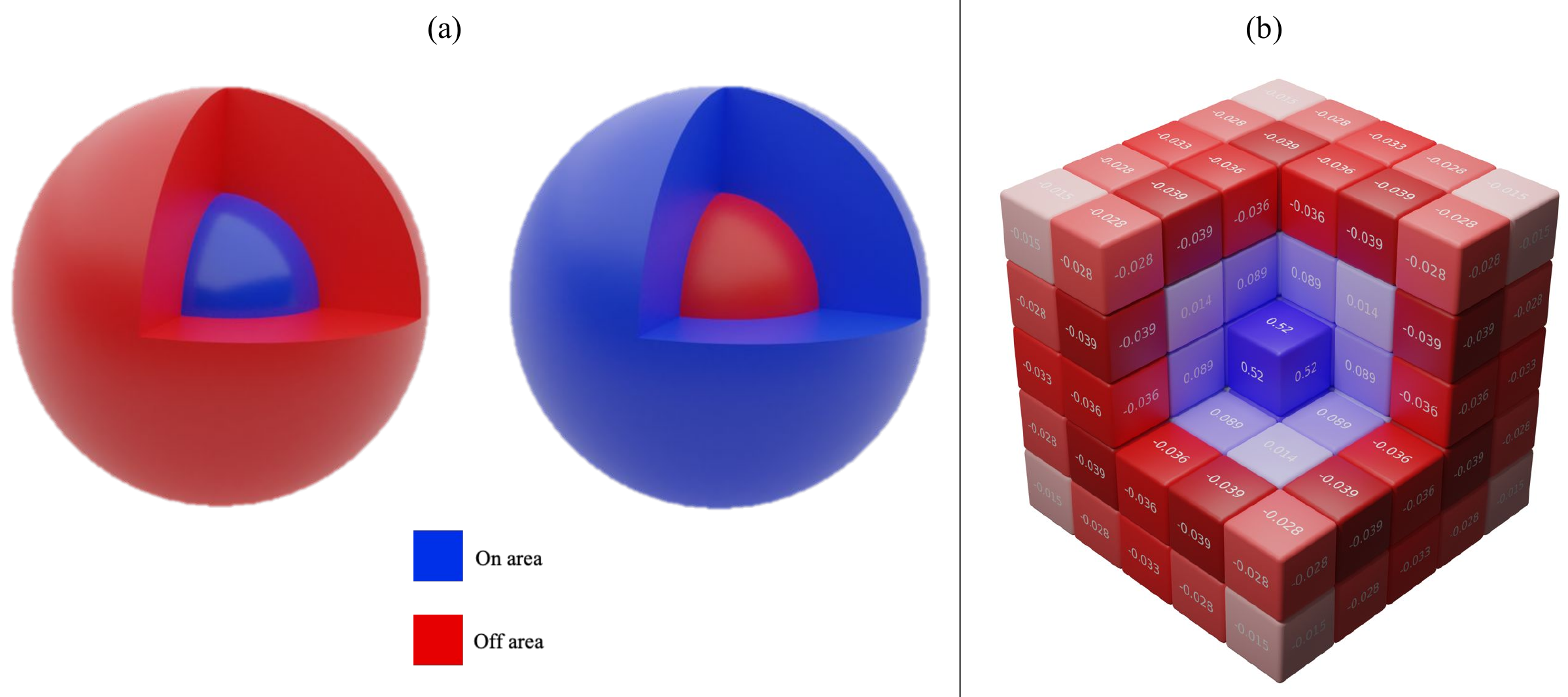}
\caption{Different geometrical representations of the IB kernels. (a) Spherical On and Off 3D center-surround receptive fields. (b) The 3D IB-On cubic kernel. The 3D-IB Off cubic kernel is complimentary; all its signs are inverted.}
\label{fig:kernelModels}
\end{figure}

\begin{figure*}[t]
\centering
\includegraphics[width=\linewidth]{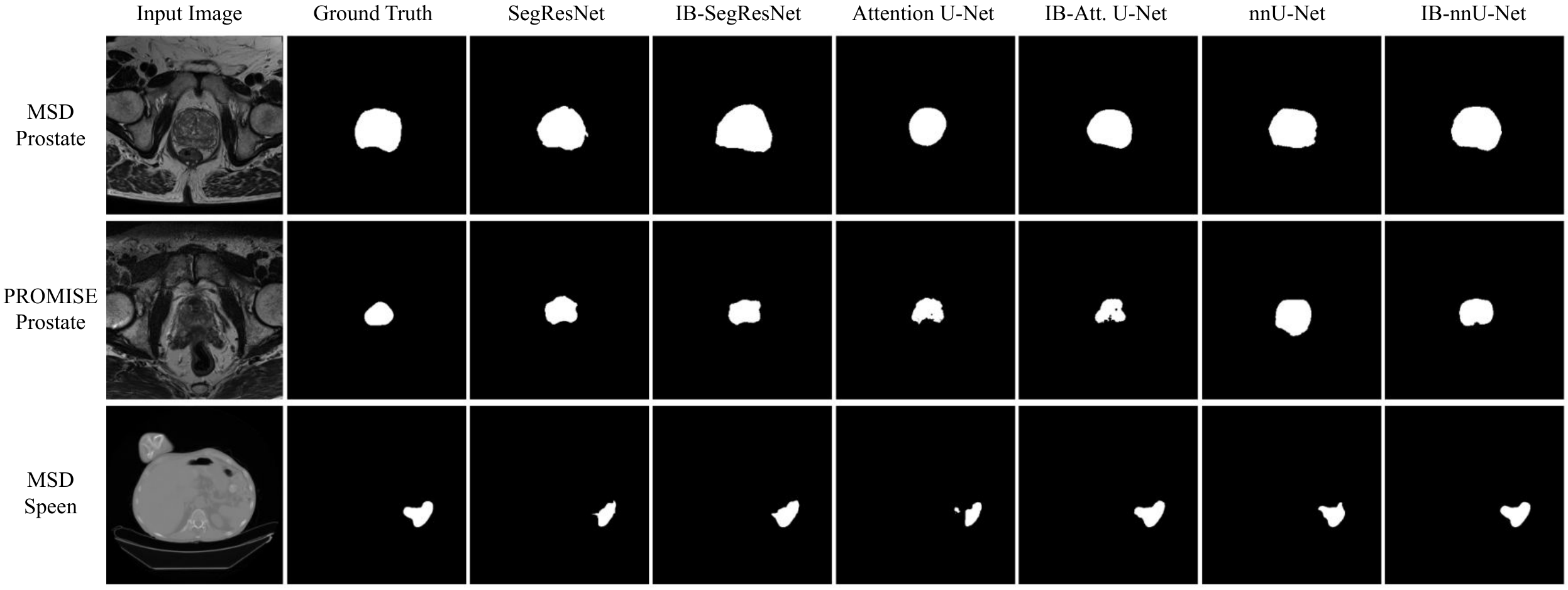}
\caption{A qualitative accuracy comparison of vanilla U-Nets and their IB extended variants on the prostate and spleen segmentation tasks. IB-nnU-Nets perform the best.}
\label{fig:SegmentationComparison}
\end{figure*}

\section{Inductive Biases in U-Net Variations}
\label{Inductive_Biases_in_U_Net_Variations}
This section first shows how we compute the two 3D IB kernels. We then discuss how to extend the architecture of U-Net variations with the two inductive-bias kernels.

\subsection{The Design of the 3D-IB Kernels}
Receptive fields in the primate retina are captured with a difference of two Gaussians (DoG)~\cite{rodieck_1965}. For 2D kernels, Equation~\ref{eq1:petrov} computes the center and surround weights~\cite{kruizinga_petkov_2000, Petkov2005ModificationsOC}:
\vspace{-2mm}
\begin{equation}
\label{eq1:petrov}
DoG_{\sigma,\gamma}(x,y) =  \frac{A_c}{\gamma^2}\, e^{-\frac{x^2+y^2}{2\gamma^2\sigma^2}} - A_s\,e^{-\frac{x^2+y^2}{2\sigma^2}}
\end{equation}
where $\gamma\,{<}\,1$ is the ratio of the radius of the center to the one of the surround, $\sigma$ is the variance of the Gaussian function, and $A_c$ and $A_s$ are the center and surround coefficients. To construct 3D-IB kernels, a naive way would be to use Equation~\ref{eq1:petrov} and repeat the kernels in the third axis. However, this approach would result in a cylindrical center-surround structure. A better way is to design a spherical structure, with excitatory center and an inhibitory surround. We therefore extend the 2D version as follows: 
\begin{equation}
\label{eq:petrov_3d}
DoG_{\sigma,\gamma}(x,y,z) =  \frac{A_c}{\gamma^3}\, e^{-\frac{x^2+y^2+z^2}{2\gamma^2\sigma^2}} - A_s\,e^{-\frac{x^2+y^2+z^2}{2\sigma^2}}
\end{equation}

We set the absolute value of the sum of the negative and positive weights to be equal to an arbitrary value $c\geq1$. This ensures the proper balance between excitation and inhibition, and the kernel weights will be large enough.
\vspace{-2mm}
\begin{equation}
[DoG_{\sigma,\gamma}(x,y,z)]^{+} dx dy dz = c,
\end{equation}
\begin{equation}
[DoG_{\sigma,\gamma}(x,y,z)]^{-} dx dy dz = -c
\end{equation}

By setting the sum of negative and positive values to the same value, the coefficients $A_c$ and $A_s$ will be equal in the infinite continuous case. To show that this result holds, we can write the DoG equation in polar coordinates: 
\begin{equation}
\int_{0}^{r_s}\!\!\!\!\int_{0}^{\pi}\!\!\!\!\int_{0}^{2\pi}\!\!\!\!\!\! r^2\,sin(\varphi)\,\left(\frac{A_c}{\gamma^3}\ e^{-\frac{r^2}{2\gamma^2\sigma^2}} - A_{s}\,e^{-\frac{r^2}{2\sigma^2}}\right) d\theta d\varphi dr
\end{equation}
where $r$ is the radius of the central sphere. We can calculate the integral when $r\,{\to}\,\infty$, and by knowing that the sum of positive and negative weights is zero, we have:
\begin{equation}
\label{eq:petrov}
\frac{2\sqrt{2}\pi^{\frac{3}{2}}}{(\frac{1}{\sigma^2})^{\frac{3}{2}}} (A_c - A_s) = 0 \ \ \Rightarrow\ \  A_c = A_s
\end{equation}
Assuming that the values of the coefficients are still close enough in the finite discrete case as well, we can approximate the variance of the Gaussians as follows:
\begin{equation}
\label{eq:sigma2}
\sigma \approx \frac{r}{\gamma}\sqrt{\frac{1-\gamma^2}{-6\ln{\gamma}}}
\end{equation}
%
%
%
%
%

\subsection{The Architecture of the Second Encoder Block}
\label{sec:TheArchitectureOfSecondEncoderBlock}
We used Equation~\ref{eq:petrov_3d} to calculate the fixed kernel weights for the On and Off (identical equation, but with signs inverted) convolutions. Note that we only needed to determine the kernel size ($k$), the ratio of the radius of the center to surround, and the sum of positive and negative values to compute the fixed kernels. Figure~\ref{fig:kernelModels} shows the $5\,{\times}\,5\,{\times}\,5$ On kernel that we use in our experiments. Here we set $\gamma=2/3$ and $r=3$. For a given input  $\chi$, we calculate the On and Off responses by convolving $\chi$ with the computed kernels separately:
\begin{equation}
\begin{array}{c@{\ }c@{\ }l}
\chi_{\rm On}[x,y,z] &=& (\chi * +DoG[r,\gamma, c])[x,y,z]\\[1mm]
\chi_{\rm Off}[x,y,z] &=& (\chi * -DoG[r,\gamma, c])[x,y,z]
\end{array}
\end{equation}

\begin{table*}[t]
\caption{Robustness of U-Net variants on the MSD-prostate dataset using DSC metric.}
\centering
\begin{tabular}{cccccc}
\toprule                  
Dataset & Subsample  & Model  & No-Noise & Gaussian Blur & Random Gaussian  \\
Name & Size & Name & & Noise, $\sigma=2$ & Noise, $\sigma=45$  \\
\midrule
MSD-prostate & $8$ & SegResNet & $0.555 \pm 0.337$ & $0.513 \pm 0.319$ & $0.540 \pm 0.225$ \\
& & IB-SegResNet & $0.560 \pm 0.231$ & $0.555 \pm 0.214$ & $0.526 \pm 0.303$ \\
& & Attention-U-Net & $0.456 \pm 0.289$ & $0.153 \pm 0.198$ & $0.520 \pm 0.308$ \\
& & IB-Att-U-Net & $0.657 \pm 0.109$ & $0.609 \pm 0.101$ & $0.627 \pm 0.112$ \\
& & nnU-Net & $0.705 \pm 0.095$ & $0.673 \pm 0.105$ & $0.704 \pm 0.098$ \\
& & IB-nnU-Net & $\mathbf{0.720 \pm 0.071}$ & $\mathbf{0.690 \pm 0.100}$ & $\mathbf{0.716 \pm 0.077}$ \\
\midrule
MSD-prostate & $16$ & SegResNet & $0.705 \pm 0.251$ & $0.464 \pm 0.301$ & $0.704 \pm 0.261$ \\
& & IB-SegResNet & $0.754 \pm 0.206$ & $0.591 \pm 0.257$ & $0.733 \pm 0.214$ \\
& & Attention-U-Net & $0.682 \pm 0.261$ & $0.144 \pm 0.266$ & $0.665 \pm 0.273$ \\
& & IB-Att-U-Net & $0.779 \pm 0.078$ & $0.742 \pm 0.105$ & $0.777 \pm 0.076$ \\
& & nnU-Net & $0.804 \pm 0.073$ & $0.758 \pm 0.095$ & $0.782 \pm 0.087$ \\
& & IB-nnU-Net & $\mathbf{0.821 \pm 0.063}$ & $\mathbf{0.789 \pm 0.085}$ & $\mathbf{0.793 \pm 0.076}$ \\
\midrule
MSD-prostate & $24$ & SegResNet & $0.805 \pm 0.181$ & $0.672 \pm 0.230$ & $0.733 \pm 0.264$ \\
& & IB-SegResNet & $0.808 \pm 0.188$ & $0.685 \pm 0.220$ & $0.790 \pm 0.193$ \\
& & Attention-U-Net & $0.816 \pm 0.119$ & $0.478 \pm 0.341$ & $0.773 \pm 0.159$ \\
& & IB-Att-U-Net & $0.819 \pm 0.113$ & $0.594 \pm 0.314$ & $0.776 \pm 0.202$ \\
& & nnU-Net & $0.823 \pm 0.065$ & $0.803 \pm 0.097$ & $0.816 \pm 0.087$ \\
& & IB-nnU-Net & $\mathbf{0.831 \pm 0.042}$ & $\mathbf{0.819 \pm 0.079}$ & $\mathbf{0.822 \pm 0.071}$ \\
\bottomrule
\end{tabular}
\label{table:Performance-Table-Metrics-Prostate}
\end{table*}

We extended a 3D U-Net segmentation variation to a 3D IB-U-Net variation by changing its second encoder block to a 3D IB encoder block with $k=5$, as shown in Figure~\ref{fig:oocsArchitecture}. Through extensive experimentation, we found out that this IB extension results in the best accuracy and robustness on the chosen small-sized datasets. For instance, initially the input to IB layers was the outputs of the max-pooling layers, but the results improved only by an insignificant margin. We also used different values for the IB parameters, such as $\gamma$ and $r$, thus creating kernels of different sizes. The use of even shaped kernels for convolution operation requires extra padding, and since this introduces artificial values, we restricted our design to kernels with odd shapes ($k=3,5,7,9$). When using a kernel size of 3 ($r=1$, $\gamma=1/2$), the performance improved, but it was it is not always guaranteed. For larger kernel sizes such as $k=7$ ($r=3$, $\gamma=3/4$) and $k=9$ ($r=4$, $\gamma=4/5$), there was a large increase in the number of network parameters and training time with an insignificant margin of improvement (less than 0.1\%) as compared to when $k=5$. 

In other scenarios, we added the IB layers in the second decoder block to maintain a degree of symmetry in the networks. However, this did not prove beneficial, as it helped for some scans but made some other volumetric segmentation results worse. In addition, we created networks where the IB layers were placed in all the encoder blocks, not just in the second encoder block. By doing this, we not only introduced more parameters, but our models started over-fitting as well. Therefore, we decided to utilize the IB kernels only in the second encoder block.

To add the On and Off IBs to the second convolution block, we first divided the convolution layers of the block into two parallel pathways, with half of the filters in the original layer in each divided convolution. This architecture mimics the so-called On and Off pathways in the retina. We added the 3D On and Off convolved inputs of the block to the activation maps of the first convolution layers of the pathways before max-pooling, such that we do not lose information. This required the use of a stride of two for the IBs. We experimentally observed that adding the IBs after max-pooling and with a stride of one leads to worse results. Finally, we concatenated the activation maps of the last layers on the pathways, resulting in the outputs with the same shapes as the outputs of the original block. 

\begin{table*}[t]
\centering
\caption{Robustness of U-Net variants on the PROMISE-12 dataset using DSC metric.}
\vspace{0.1in}
\begin{tabular}{ ccccc }
\toprule                  
Subsample  & Model  & No-Noise & Gaussian Blur & Random Gaussian  \\
Size & Name & & Noise, $\sigma=2$ & Noise, $\sigma=45$  \\
\midrule
$8$ & SegResNet & $0.559 \pm 0.267$ & $0.392 \pm 0.311$ & $0.510 \pm 0.326$ \\
 & IB-SegResNet & $0.624 \pm 0.214$ & $0.495 \pm 0.246$ & $0.639 \pm 0.189$ \\
 & Attention-U-Net & $0.512 \pm 0.325$ & $0.079 \pm 0.123$ & $0.508 \pm 0.317$ \\
 & IB-Att-U-Net & $0.674 \pm 0.104$ & $0.623 \pm 0.199$ & $0.662 \pm 0.112$ \\
 & nnU-Net & $0.722 \pm 0.136$ & $0.704 \pm 0.175$ & $0.721 \pm 0.146$ \\
 & IB-nnU-Net & $\mathbf{0.735 \pm 0.127}$ & $\mathbf{0.713 \pm 0.159}$ & $\mathbf{0.719 \pm 0.135}$ \\
\midrule
$16$ & SegResNet & $0.618 \pm 0.267$ & $0.603 \pm 0.301$ & $0.611 \pm 0.268$ \\
 & IB-SegResNet & $0.705 \pm 0.213$ & $0.650 \pm 0.297$ & $0.678 \pm 0.224$ \\
 & Attention-U-Net & $0.582 \pm 0.298$ & $0.241 \pm 0.304$ & $0.589 \pm 0.273$ \\
 & IB-Att-U-Net & $0.622 \pm 0.251$ & $0.461 \pm 0.349$ & $0.617 \pm 0.261$ \\
 & nnU-Net & $0.759 \pm 0.155$ & $0.740 \pm 0.169$ & $\mathbf{0.751 \pm 0.160}$ \\
 & IB-nnU-Net & $\mathbf{0.779 \pm 0.146}$ & $\mathbf{0.743 \pm 0.158}$ & $0.738 \pm 0.175$ \\
\midrule
$24$ & SegResNet & $0.645 \pm 0.234$ & $0.532 \pm 0.244$ & $0.594 \pm 0.296$ \\
 & IB-SegResNet & $0.768 \pm 0.130$ & $0.596 \pm 0.237$ & $0.723 \pm 0.239$ \\
 & Attention-U-Net & $0.690 \pm 0.234$ & $0.449 \pm 0.302$ & $0.640 \pm 0.245$ \\
 & IB-Att-U-Net & $0.737 \pm 0.162$ & $0.487 \pm 0.307$ & $0.645 \pm 0.288$ \\
 & nnU-Net & $0.803 \pm 0.104$ & $0.794 \pm 0.125$ & $0.800 \pm 0.124$ \\
 & IB-nnU-Net & $\mathbf{0.811 \pm 0.095}$ & $\mathbf{0.800 \pm 0.120}$ & $\mathbf{0.802 \pm 0.108}$ \\
\bottomrule
\end{tabular}
\label{table:Performance-Table-Metrics-PROMISE-12}
\end{table*}

\vspace*{-2mm}
\section{Materials and Methods}
\label{Materials_and_Methods}

\subsection{U-Net Variations Used in Our Experiments}
\label{Networks}
To explore the merit of IBs in U-Nets, we first examined the accuracy and robustness of all variations mentioned in Section \ref{related_work} and their extension. Attention-U-Nets and SegResNets introduced novel multi-view and multi-scale functions, which enhanced the receptive fields of the networks and reduced bottlenecks during computation, essentially shortening the time required to converge. Furthermore, these architectures showed excellent performances on the public dataset. Therefore, we proceeded to further benchmark these two architectures in our framework. However, we excluded the V-Net from this paper, as the SegResNet has a similar structure as that of the former model. Moreover, the SegResNet improves on the V-Net by introducing a VAE branch. 
While all variations benefited from the IB extension, nnU-Nets, Attention-U-Nets and SegResNets performed best. nnU-Nets, Attention-U-Nets and SegResNets were extended with 3D-IB filters of kernel sizes $k=5$. The resulting new networks are called as follows: IB-nnU-Net, IB-Att-U-Net  and IB-SegResNet respectively. 

\subsection{Datasets}
\label{Datasets}
In all our experiments concerning the prostate organ, we utilized two MRI datasets from two different public sources: PROMISE-12 and MSD-prostate. 
\paragraph{PROMISE-12:} The PROMISE-12 dataset offers 80 (training=50, testing=30) volumes of transversal T2-weighted MRIs of the prostate. This dataset was selected as it is a compilation of scans collected from multiple centres and vendors with different acquisition protocols. Due to this fact, there is a heterogeneity in the voxel-spacings and the slice thickness of the volumes. 
\begin{table*}[t]
\centering
\caption{Robustness of U-Net variants on the MSD-spleen dataset using DSC metric.}
\begin{tabular}{ ccccc }
\toprule                  
Subsample  & Model  & No-Noise & Gaussian Blur & Random Gaussian  \\
Size & Name & & Noise, $\sigma=2$ & Noise, $\sigma=45$  \\
\midrule
$8$ & SegResNet & $0.601 \pm 0.311$ & $0.605 \pm 0.304$ & $0.593 \pm 0.294$ \\
 & IB-SegResNet & $0.621 \pm 0.271$ & $0.618 \pm 0.285$ & $0.616 \pm 0.274$ \\
 & Attention-U-Net & $0.615 \pm 0.293$ & $0.603 \pm 0.303$ & $0.611 \pm 0.300$ \\
 & IB-Att-U-Net & $0.630 \pm 0.235$ & $0.612 \pm 0.289$ & $0.615 \pm 0.275$ \\
 & nnU-Net & $0.652 \pm 0.247$ & $0.628 \pm 0.244$ & $0.650 \pm 0.236$ \\
 & IB-nnU-Net & $\mathbf{0.665 \pm 0.233}$ & $\mathbf{0.631 \pm 0.241}$ & $\mathbf{0.659 \pm 0.234}$ \\
\midrule
$16$ & SegResNet & $0.704 \pm 0.197$ & $0.661 \pm 0.211$ & $0.679 \pm 0.201$ \\
 & IB-SegResNet & $0.715 \pm 0.183$ & $0.704 \pm 0.190$ & $0.711 \pm 0.178$ \\
 & Attention-U-Net & $0.728 \pm 0.171$ & $0.719 \pm 0.180$ & $0.720 \pm 0.184$ \\
 & IB-Att-U-Net & $0.739 \pm 0.168$ & $0.722 \pm 0.181$ & $0.734 \pm 0.169$ \\
 & nnU-Net & $\mathbf{0.754 \pm 0.152}$ & $0.730 \pm 0.168$ & $\mathbf{0.746 \pm 0.149}$ \\
 & IB-nnU-Net & $0.751 \pm 0.154$ & $\mathbf{0.732 \pm 0.166}$ & $0.741 \pm 0.150$ \\
\midrule
$24$ & SegResNet & $0.818 \pm 0.118$ & $0.776 \pm 0.146$ & $0.765 \pm 0.148$ \\
 & IB-SegResNet & $0.819 \pm 0.109$ & $0.799 \pm 0.121$ & $0.786 \pm 0.135$ \\
 & Attention-U-Net & $0.841 \pm 0.088$ & $0.824 \pm 0.114$ & $0.817 \pm 0.122$ \\
 & IB-Att-U-Net & $0.843 \pm 0.080$ & $0.833 \pm 0.091$ & $0.829 \pm 0.106$ \\
 & nnU-Net & $0.868 \pm 0.067$ & $0.840 \pm 0.102$ & $\mathbf{0.852 \pm 0.093}$ \\
 & IB-nnU-Net & $\mathbf{0.872 \pm 0.067}$ & $\mathbf{0.842 \pm 0.100}$  & $0.851 \pm 0.093$ \\
\bottomrule
\end{tabular}
\label{table:Performance-Table-Metrics-Spleen}
\vspace*{-2mm}
\end{table*}

\paragraph{MSD-prostate:} The MSD-prostate dataset comprises of 48 (training=32, testing=16) multimodal (T2, ADC) 3D MRI samples. We only considered the T2-weighted modality as they provide most of the necessary anatomical information, whereas the ADC modality is often considered for tumour characterization/segmentation. Furthermore, the ground-truth labels of the original MSD-prostate dataset are separated into annotations of the central prostate gland and the peripheral zone. We combined the two regions into one label such that all U-Net variations could learn a binary image segmentation. Similar to the PROMISE-12, the MSD-prostate dataset was selected as it has a significant amount of inter-subject variability.

\paragraph{MSD-spleen:}
The spleen dataset consists of 61 (training=41, testing=20) monomodal 3D CT samples. We wanted to work with this dataset because of two reasons; first, it has a small sample size, and second, we wanted to check the potential of the IB kernels when handling CT images with large variations in their field-of-view.

\subsection{Implementation of our framework}
Other than extending the networks with the IB kernels, we did not make any alterations to the original networks, including the convolution and normalization layers~\cite{ioffe_batch_norm_2015}, activation functions, and dropout layers and probabilities. All the necessary source code for the framework was implemented in PyTorch~\cite{pytorch_2019_NIPS}. 

The volumes in the three datasets have varying voxel spacings, and CNNs disregard this information while operating on the voxel grids. To handle this heterogeneity in the datasets, all the images and their corresponding ground truth labels were resampled to the target voxel spacings of their respective datasets: $0.6125 \times 0.6125 \times 2.2 mm^3$ for PROMISE-12, $0.625 \times 0.625 \times 3.6 mm^3$ for the MSD-prostate and $0.79 \times 0.79 \times 1.6 mm^3$ for MSD-spleen. We employed tri-linear and nearest-neighbour interpolation to resample the images and annotated labels, respectively. In most biomedical cases, we cannot fit the complete image into the GPU due to the large sizes. For that reason, 3D patches of foreground and background were randomly sampled with equal probabilities. 
The shape of the sampled patches were: $192\,{\times}\, 192\,{\times}\,16$ voxels for the PROMISE-12 dataset, $288\,{\times}\, 288\,{\times}\,16$ for the MSD-prostate set, and $192\,{\times}\, 192\,{\times}\,96$ for the spleen dataset. 

The intensity values of the MR scans were linearly scaled using a standard (z-score) normalization such that the mean value and standard deviation of the transformed data equalled to 0 and 1, respectively. As for CT images, the first 0.5 and 99.5 percentiles of the foreground voxels were clipped, and then normalized using the global foreground mean and standard deviation. To synthesize additional data during training, the data pipeline carried out augmentations, such as elastic deformations, gamma correction, random Gaussian noise, Gaussian blur, rotations, and scaling. Throughout all the experiments, we used the Adam optimization algorithm with an initial learning rate of 1e-04 that decays using a polynomial scheduler ~\cite{segresnet2018} for 500 epochs. We used L2 regularization on the convolution kernel parameters with a weight of 1e-05. The gradient updates were evaluated with a mini-batch size of two for all the datasets. 

\paragraph{Loss functions:} We utilized a combination of two-loss functions; Binary Cross-Entropy loss (L\textsubscript{BCE}) and Sørensen-Dice Loss (L\textsubscript{dice}) \cite{sudre_generalised_2017} together make BCE-dice loss (L\textsubscript{BCE-dice}) \cite{vnet_2016} to train the models. This compound loss is defined over all semantic classes and is less sensitive to class imbalance, as demonstrated experimentally in \citet{MA2021102035}. The equations for BCE-dice loss are given as follows: 
\begin{equation}
\begin{array}{l@{\ }l}
L\textsubscript{BCE} & =  - \frac{1}{N} \cdot \sum_{i=1}^N g_i\cdot log(p_i) + (1-g_i)\cdot log(1-p_i)
\end{array}
\end{equation}
\vspace{-2mm}
\begin{equation}
\begin{array}{l@{\ }l}
L\textsubscript{dice} & = 1 - \frac{2*\sum_{i=1}^N p_i g_i}{\sum_{i=1}^N p_i^2 + \sum_{i=1}^N g_i^2}
\end{array}
\end{equation}
\vspace{-2mm}
\begin{equation}
\begin{array}{l@{\ }l}
L\textsubscript{BCE-dice} &= L\textsubscript{BCE} + L\textsubscript{dice}
\end{array}
\end{equation}
$N$ is the number of voxels, $p_i$ and $g_i$ denote the prediction and ground-truth label at voxel $i$ respectively, for an image.

\begin{table*}[t]
\centering
\caption{Accuracy of U-Net variants and their IB extensions on the full datasets without any artifacts.}
\begin{tabular}{cccc}
\toprule                  
Model Name & MSD-prostate ($32$) & PROMISE-12 ($50$) & Spleen dataset ($41$) \\
\midrule
SegResNet               & $0.841 \pm 0.085$ &   $0.885 \pm 0.035$ &   $0.901 \pm 0.083$ \\
IB-SegResNet       & $\mathbf{0.844 \pm 0.078}$ &   $\mathbf{0.888 \pm 0.033}$ &  $\mathbf{0.902 \pm 0.080}$ \\
\midrule
Attention-U-Net         & $0.835 \pm 0.075$ &   $0.891 \pm 0.033$ &   $0.938 \pm 0.062$ \\
IB-Att-U-Net & $\mathbf{0.845 \pm 0.066}$ &   $\mathbf{0.892 \pm 0.032}$ &   $\mathbf{0.941 \pm 0.058}$ \\
\midrule
nnU-Net & $0.882 \pm 0.118$ & $0.890 \pm 0.101$ & $0.966 \pm 0.044$ \\
IB-nnU-Net & $\mathbf{0.895 \pm 0.042}$ & $\mathbf{0.902 \pm 0.036}$ & $\mathbf{0.970 \pm 0.017}$ \\
\bottomrule
\end{tabular}
\label{table:Performance-Table-Full-Datasets}
\vspace{-0.1in}
\end{table*}

In the validation phase, neither patch sampling was done nor any augmentations were applied, rather only the appropriate normalization techniques were applied to the full image. Next, to predict the label, a sliding window technique was used with an overlap of $1/2$ in each dimension, and the window shape was the set to the training patch size. The state values of the model configuration with the highest metric value was saved and later used for inference. The models were evaluated based on the DSC metric (Dice score), as it exhibits a low overall bias in 3D medical image segmentation tasks ~\cite{taha_metrics_2015, maier2022metrics}. The Dice score is defined in equation \ref{eq:dsc}, where $\epsilon$ is the smoothing factor with a value of $10^-6$.


\begin{equation}
\label{eq:dsc}
\begin{array}{l@{\ }l}
\text{DSC} & = \frac{2*\sum_{i=1}^N p_i g_i + \epsilon}{\sum_{i=1}^N p_i^2 + \sum_{i=1}^N g_i^2 + \epsilon}
\end{array}
\end{equation}


Once the training was finished, first the voxel-wise semantic predictions were generated using the best network parameters, and then the labels were resampled to their original spacings. Next, we retrieved the desired region by taking the largest connected component (apart from the background) of the resultant resampled volume. The inference step ended after the final metric values were calculated, and the detailed results were saved.


\section{Experiments and Results}
The proposed 3D IB filters are modular and independent of the network's architecture, allowing them to be easily deployed to image segmentation tasks. Unfortunately, the ground-truth labels for the testing images are not publicly available for any of the datasets mentioned in Section \ref{Datasets}. Consequently, to analyze the capabilities of the networks, we conducted two kinds of CV using the 50 MR volumes, 32 MR images and 41 CT scans from the PROMISE-12, MSD-prostate, and MSD-spleen datasets, respectively. First, we implemented 8-fold CV by randomly sampling three subsets each from the datasets. The sample size was chosen such that it would be a factor of $8$, hence, the number of volumes in each sub-sample was $8$, $16$ and $24$ respectively. This made sure that we could investigate the capabilities of our IB extensions on small-sized datasets. Second, we investigated the models for all available volumes in the three datasets for completeness using 5-fold CV. For the nnU-Net framework, we extended the U-Net of the nnU-Net with the IB kernels, and then followed the steps detailed in \citet{nnUNet_2020}.  We ran our experiments on an NVIDIA Titan RTXs with 24 GB memory. The CV method for a single network on one dataset was completed in four and six days with our framework and when using the nnU-Net framework, respectively. We acknowledge that training models for a long period could possibly negatively impact the environment, but this can be mitigated in future works by training the best model presented in this paper with a 5-fold CV method. 

Initially, we examined the accuracy and robustness of the SegResNets, Attention-U-Nets, nnU-Nets and their IB extensions on the original volumes. Figure~\ref{fig:SegmentationComparison} illustrates a few of the qualitative examples by comparing the predictions of the U-Net variations and their IB extensions, and it is evident that the IB extensions are more accurate than the original U-Nets. The IB-nnU-Net attained the highest result. In the next stage, we corrupted the original volumes by adding two types of noise: Gaussian blur and random Gaussian noise. We used the Torchio package \cite{perez-garcia_torchio_2021} to create the noisy datasets. 

\subsection{Robustness Evaluation on Small Subsets}  
Tables~\ref{table:Performance-Table-Metrics-Prostate},~\ref{table:Performance-Table-Metrics-PROMISE-12} and~\ref{table:Performance-Table-Metrics-Spleen} summarise our accuracy and robustness results for SegResNets, Attention-U-Nets, nnU-Nets and their IB extensions. The best performance is emphasized in boldface. The key takeaway is that IB extensions are more performant. As demonstrated by Table~\ref{table:Performance-Table-Metrics-Prostate}, the IB extensions always have superior performance on the MSD-prostate dataset. Additionally, in  Tables~\ref{table:Performance-Table-Metrics-PROMISE-12} and~\ref{table:Performance-Table-Metrics-Spleen}, a similar observation is also true for the PROMISE-12 and the spleen dataset, with a few exceptions: the nnU-Net for 16 and 24 subsamples. In these cases, the lower performance of the IB extension could be attributed to the presence of outliers. Please note, however, that the variance in these two cases is relatively high. Thus, the results for the IB extensions are still within the same range as the ones for original nnU-Net. In general, the variance of the vanilla U-Net variants is larger in these tables than the one of their IB extensions.

\subsection{Robustness Evaluation on Full Dataset}
For completeness, in Table~\ref{table:Performance-Table-Full-Datasets} we provide the DSC-accuracy results of SegResNets, Attention-U-Nets, nnU-Nets and their IB-extensions, for all available volumes in the three datasets. Similar to Tables~\ref{table:Performance-Table-Metrics-Prostate},~\ref{table:Performance-Table-Metrics-PROMISE-12} and~\ref{table:Performance-Table-Metrics-Spleen}, the IB-extensions are performing better for full datasets. However, the performance improvement is not that significant any more. In other words, the smaller the dataset, the more rectifying is the addition of IB to the U-Nets. This is true in general for machine learning (ML): the smaller the dataset, the more important is the ML algorithm.

\subsection{Cylindrical versus Spherical IB kernels}
The extension of the IB kernel from 2D to 3D is non-trivial. If we assume that the 3D IB kernel is simply the repeated stacking of the 2D kernel on top of one another (along the z-axis, with each slice having the same weights), then this leads to a cylindrical kernel. Table~\ref{table:Cylindrical_Spherical_comparison} shows the performance of the spherical kernel against the cylindrical kernels for the Attention U-Net. As one can see, our spherical kernel outperforms other variants, while the cylindrical versions are inferior compared to the original Attention U-Net. The cylindrical kernels make mistakes when considering 3D contours; therefore, it is crucial to make modifications. Moreover, \citet{pmlr-v139-babaiee21a} used natural images to evaluate the performance of the 2D IB kernels, and the kernels that work for 2D may not necessarily be suitable in the 3D setting. The properties of 3D medical images, such as the voxel spacings, matrix directions, origins etc., produce significant artifacts that the cylindrical kernels cannot handle well.

\begin{table}[t]
\centering
\caption{Comparison of robustness among Attention U-Net, cylindrical IB-attention U-Nets ($k=3$ and $k=5$), and spherical IB-attention U-Net ($k=5$) on subsample size 8 of the PROMISE-12 dataset using the DSC metric. To be added in the final version.}
\label{table:Cylindrical_Spherical_comparison}
\vspace{0.5mm}
\begin{tabular}{ cc }
\toprule                  
Model Name & DSC ($\uparrow$) \\
\midrule
Attention-U-Net & $0.512 \pm 0.325$ \\
Cyl IB-Att. U-Net (k3) & $0.438 \pm 0.358$ \\
Cyl IB-Att. U-Net (k5) & $0.445 \pm 0.336$ \\
IB-Att. U-Net (k5) & $\mathbf{0.674 \pm 0.104}$ \\
\bottomrule
\end{tabular}
\end{table}

\begin{table}[t]
\centering
\caption{Comparison of robustness amongst nnU-Net and its IB-extended U-Nets (k3, k5, k7 and k9) on the full PROMISE-12 dataset using the DSC metric. We will add it in the final version.}
\label{table:Performance_hyperparameters}
\vspace{0.5mm}
\begin{tabular}{ cc }
\toprule                  
Model Name & DSC ($\uparrow$) \\
\midrule
nnU-Net & $0.890 \pm 0.101$ \\
IB-nnU-Net ($k=3$, $r=1$, $\gamma=1/2$) & $0.887 \pm 0.086$ \\
IB-nnU-Net ($k=5$, $r=2$, $\gamma=2/3$) & $\mathbf{0.902 \pm 0.036}$ \\
IB-nnU-Net ($k=7$, $r=3$, $\gamma=3/4$) & $0.898 \pm 0.055$ \\
IB-nnU-Net ($k=9$, $r=4$, $\gamma=4/5$) & $0.895 \pm 0.058$ \\
\bottomrule
\end{tabular}
\vspace{-4mm}
\end{table}

\begin{table}[t]
\centering
\caption{Accuracy of nnU-Net and IB-nnU-Net on the full datasets without any artifacts using the HD95 (in mm) metric. To be added.}
\label{table:HD95_metric}
\vspace{0.5mm}
\begin{tabular}{ ccc }
\toprule                  
Dataset Name & Model Name & HD95 in mm ($\downarrow$) \\
\midrule
MSD-prostate & nnU-Net & $ 15.078 \pm 75.203$ \\
 & IB-nnU-Net & $\mathbf{1.738 \pm 1.122}$ \\
\midrule
PROMISE-12 & nnU-Net & $ 9.735 \pm 59.565$ \\
 & IB-nnU-Net & $\mathbf{1.237 \pm 0.556}$ \\
\midrule
MSD-spleen & nnU-Net & $ 1.640 \pm 4.708$ \\
 & IB-nnU-Net & $\mathbf{1.189 \pm 1.914}$ \\
\bottomrule
\end{tabular}
\end{table}

\subsection{Different parameters of IB kernels}
The section on the architecture of the second encoder block discusses the performances of the various IB-extended U-Net architectures, such as U-Nets with multiple IB blocks; however, their performances were not satisfactory. We also conducted experiments with different $k$, $r$, and $\gamma$, and the quantitative results are given in Table ~\ref{table:Performance_hyperparameters}. It is clear that the IB-nnU-Net variant with parameter values of $k=5$, $r=2$, and $\gamma=2/3$ is the best performing version

\section{Discussion and Conclusion}
\label{Discussion_and_Conclusion}
We introduced two new 3D kernels, inspired by the on and off center-surround pathways originating in the vertebrate retina, and we proposed a procedure to compute the kernel weights. We showed how to extend the second encoder block of U-Net variations, using these precomputed kernels as inductive biases, helping to scrutinize and delineate anatomical structures of 3D images with increased accuracy. Although our proposed 3D-IB-encoder block slightly increases the number of training parameters of the networks, the modification steps are effortless, and it requires minimal computational overhead. 

The IB kernels can help in scenarios where the dataset has at least two outliers (very different samples). In this case, the other models have difficulties in segmenting the region of interest (ROI). This is evident in Figures~\ref{fig:attention_maps} and ~\ref{fig:SegmentationComparison}, as well as in Table~\ref{table:Performance-Table-Full-Datasets}, where the standard deviation of the best performing IB-nnU-Net is smaller than that of the second place nnU-Net. The marginal improvement of performances in Table~\ref{table:Performance-Table-Metrics-Spleen} could be due to sampling very similar volumes from the spleen task without any outliers. In the full MSD-spleen dataset, the ROI is so large that when the model prediction is incorrect by a significant margin, the DSC metric may not capture this distinction accurately. Therefore, in Table~\ref{table:HD95_metric} we have listed the segmentation performances of the nnU-Net and IB-nnU-Net, respectively, using a surface distance metric called the 95th percentile Hausdorff distance (HD95).

Robustness to distribution shifts is a crucial attribute of any deep neural network, and it is of even higher importance for networks used in critical domains like bio-medical tasks. Our experiments on multiple organ segmentation tasks with our baseline networks and their IB extensions show notable enhancements in the performance of the IB-equipped networks. Furthermore, our robustness experiments demonstrate the superior performance of IB extended networks. Figure \ref{fig:attention_maps} shows that the IB extension improves a network's focus on critical structures. In conclusion, the proposed method could make the 3D-IB kernels a favourable approach to achieve accurate results in 3D medical image segmentation tasks with small sized datasets.

\section{Acknowledgements}
This research was funded in part by the Austrian Science Fund (FWF): I 4718, and the Federal Ministry of Education and Research (BMBF) Germany, under the frame of ERA PerMed. Z. B. was supported by the Doctoral College Resilient Embedded Systems, which is run jointly by the TU Wien's Faculty of Informatics and the UAS Technikum Wien.


\section{Code and Data Availability} 
All code and data are included in https://github.com/Shrajan/IB\_U\_Nets.

\bibliography{aaai22}

\end{document}